\documentclass[conference,compsoc]{IEEEtran}

\usepackage{graphicx} 

\usepackage[nointegrals]{wasysym} 
\usepackage[normalem]{ulem} 
\usepackage[style=french]{csquotes} 
\usepackage[most]{tcolorbox} 
\usepackage{cleveref}
\usepackage[table]{xcolor} 
\usepackage{makecell} 
\usepackage{multirow}
\usepackage{booktabs} 
\usepackage{subcaption} 
\usepackage[T1]{fontenc} 
\usepackage{url}
\usepackage{cite}
\usepackage[numbers, sort&compress]{natbib}

\usepackage{enumitem}
\setlistdepth{9}

\setlist[itemize,1]{label=$\bullet$}
\setlist[itemize,2]{label=$-$}
\setlist[itemize,3]{label=$*$}
\setlist[itemize,4]{label=$\bullet$}
\setlist[itemize,5]{label=$-$}
\setlist[itemize,6]{label=$*$}
\setlist[itemize,7]{label=$\bullet$}
\setlist[itemize,8]{label=$-$}
\setlist[itemize,9]{label=$*$}

\newcommand{\new}[1]{#1}

\newcommand\HPOone{\textit{WA-M-1}}
\newcommand\HPOtwo{\textit{SH-M-1}}
\newcommand\HPOthree{\textit{SH-U-1}}
\newcommand\HPOfive{\textit{WA-M-2}}
\newcommand\HPOsix{\textit{SH-U-2}}
\newcommand\HPOeight{\textit{WA-U-1}}
\newcommand\HPOnine{\textit{VPS-U-1}}
\newcommand\HPOten{\textit{WA-M-3}}
\newcommand\HPOeleven{\textit{VPS-U-2}}
\newcommand\HPOtwelve{\textit{VPS-MU-1}}
\newcommand\HPOthirtn{\textit{VPS-M-1}}
\newcommand\HPOfourtn{\textit{VPS-U-3}}
\newcommand\HPOfiftn{\textit{SH-M-2}}

\newcommand\HPOseventn{\textit{SH-M-3}}
\newcommand\HPOeightn{\textit{SH-U-3}}
\newcommand\HPOtwenty{\textit{SH-MU-1}}
\newcommand\HPOtwtyone{\textit{VPS-U-4}}
\newcommand\HPOtwtytwo{\textit{SH-M-4}}
\newcommand\HPOtwtythree{\textit{VPS-MU-2}}
\newcommand\HPOtwtyfour{\textit{SH-M-7}}
\newcommand\HPOtwtyfive{\textit{VPS-U-5}} 
\newcommand\HPOtwtysix{\textit{SH-U-4}}
\newcommand\HPOtwtyseven{\textit{SH-M-5}}
\newcommand\HPOtwtyeight{\textit{SH-M-6}}

\begin{document}
\title{Behind the Curtain: How Shared Hosting Providers Respond to Vulnerability Notifications}

\author{\IEEEauthorblockN{Giada Stivala,  Rafael Mrowczynski, Maria Hellenthal, Giancarlo Pellegrino}
\IEEEauthorblockA{CISPA Helmholtz Center for Information Security\\
Saarbr\"{u}cken, Germany\\
stivala.gm@proton.me, \{mrowczynski, hellenthal, pellegrino\}@cispa.de}
}

\maketitle

\begin{abstract}
Large-scale vulnerability notifications (VNs) can help hosting provider organizations (HPOs) identify and remediate security vulnerabilities that attackers can exploit in data breaches or phishing campaigns. Previous VN studies have primarily focused on factors under the control of reporters, such as sender reputation, email formatting, and communication channels. Despite these efforts, remediation rates for vulnerability notifications continue to remain consistently low. 
This paper presents the first in-depth study of how HPOs process vulnerability notifications internally and what organizational and operational factors influence VN effectiveness. We examine the problem from a different perspective to provide the first detailed understanding of the reasons behind persistently low remediation rates. Instead of manipulating parameters of VN campaigns, we interview hosting providers directly, investigating how they handle vulnerability notifications and what factors may influence VN effectiveness, such as VN awareness and reachability, HPOs’ service models, and perceived security risks.

We conducted semi-structured interviews with 24 HPOs across shared hosting and web development services, representing varied company sizes and operator roles. Our findings reveal practical insights on VN processing and abuse workflows. 
While some providers remain hard to reach due to complex infrastructures, most report routinely handling VNs. However, limited remediation often stems from strict responsibility boundaries, where web application issues are seen as the customer's domain. Low hosting fees and high volumes of daily compromises further discourage both proactive and reactive measures.
Our findings show that HPOs blame negligent website owners, and prior works on website owners confirms they often undervalue their sites or lack security know-how. This misalignment raises further concerns about the efficacy of current VN approaches and whether they can reliably prompt remedial action under the existing operational model.

\end{abstract}

\IEEEpeerreviewmaketitle

\section{Introduction}

The security of the Web ecosystem is critical, as vulnerabilities in web services can expose sensitive information and systems to cyberattacks, posing risks not only to website owners but also to hosting companies and other Internet users.
A common approach to addressing web vulnerabilities are vulnerability notifications, where the (presumably) responsible party is informed after detection--a practice widely used in large-scale studies by the academic security community.
However, identifying and reaching the appropriate contact poses significant challenges, undermining both issue awareness and remediation decisions.

Previous research has extensively explored ways to improve remediation rates for vulnerability notifications by modifying specific properties of VNs, such as email features~\cite{stock2018didn, li2016you} and sender's attributes~\cite{cetin2016understanding, stock2018didn}, the nature and severity of the vulnerability~\cite{stock2018didn, utz2023comparing, maass2021effective}, the communication channel~\cite{cetin2017make, stock2018didn, maas2021snail, maass2021effective, stock2016hey}, or the role of the recipient~\cite{stover2023website, cetin2017make, stock2016hey, canali2013role}.
Despite these efforts, the response and remediation rates for notified web vulnerabilities still average between 20 to 30\%~\cite{li2016you, stock2016hey, cetin2017make, stock2018didn}, and no effective alternatives have yet been identified (e.g., still no better channel than WHOIS~\cite{stock2016hey, cetin2017make, poteat2021you}).

While studies with website owners offered some understanding of their VN evaluation criteria~\cite{hennig2022standing, maass2021effective}, website owners are usually not the recipients of such notifications, as their contacts are not easily available at a large scale.
Other studies have explored the causes of non-remediation through interviews with website and system operators~\cite{bondar2023internet, stock2018didn}, revealing a complex landscape. Operators report of multiple undiscovered security misconfigurations that could lead to incidents~\cite{dietrich2018investigating}, which underscores the importance of VNs for raising awareness. However, awareness alone does not guarantee remediation~\cite{stock2018didn}, suggesting that factors beyond VN characteristics influence these decisions. Further research examined whether technical or organizational factors impacted remediation but found no single factor to be more significant than others~\cite{bondar2023internet}.
As a result, we still lack a clear understanding of what happens after a vulnerability notification is received, as none of the previous works has investigated the internal processes of the receiving entities.

To address this gap, we turn our focus to hosting provider organizations, which frequently receive vulnerability notifications due to their accessible contact information and capacity to intervene in affected customer instances. To our best knowledge, this is the first study to deeply investigate how HPOs internally receive, process, and respond to vulnerability notifications, providing a previously missing perspective on these workflows. Our first objective is to examine how these notifications are handled from the receiver’s side, offering insights into the features and processes that researchers have so far only hypothesized. Our second, and main, objective is to identify the organizational and operational factors, beyond the characteristics of the notifications themselves, that influence IT operators at HPOs in their decision to remediate.
To answer our research questions, we conducted a qualitative study based on semi-structured interviews with 24 IT operators working within HPOs, covering shared hosting providers and web agencies of varying sizes and service models.
Rather than focusing on the design of notification campaigns, our study explores how HPOs receive, interpret, and act on vulnerability notifications in practice, from the inside. 

Our findings show that while most providers are aware of and routinely handle VNs, remediation is often deprioritized due to clearly defined service boundaries, operational cost considerations, and the perception that responsibility lies with customers. 
Providers generally take responsibility for managing infrastructure and removing abusive content such as phishing, malware, or illegal material. However, they see the underlying causes of such abuse, lying in the security of web applications, as the responsibility of customers or their web agencies.
Moreover, we find that organizational flexibility, rather than rigid processes, characterizes how notifications are handled, with decisions frequently left to individual operators. Although reachability issues persist in some multi-layered hosting setups, most providers can be contacted through established channels. Ultimately, our results suggest that the limited effectiveness of VNs is not due to technical or procedural obstacles, but to a misalignment between the goals of security researchers and the business logic of hosting providers. This disconnect is deepened by providers’ frustration with what they describe as widespread negligence among website owners, raising doubts about the viability of remediation within the current shared hosting ecosystem.

\section{Background and Methods}
Our investigation was driven by the observed low remediation rates to web vulnerability notifications and our desire to uncover the roadblocks that prevent addressing these problems.
Against this background, we investigate: 

\textbf{RQ1}: \textit{How are vulnerability notifications received and processed within HPOs?}

\textbf{RQ2}: \textit{What are the characteristics of HPOs (their internal factors) that influence their VN-handling and the respective remediation processes?}

To answer these research questions, we first conducted a content analysis of HPO websites. This analysis provided a map of the research field and informed the development of our sampling strategy. Using this strategy, we then carried out a qualitative study based on semi-structured interviews with individuals owning or working for small- to large-sized hosting provider companies.
The interviews focused on VN remediation both in detail and from a broader perspective, while allowing participants to share their own viewpoints.

Data analysis and interpretation of the interview study were carried out using a combination of Qualitative Content Analysis~\cite{Mayring.2000, Schreier.2012, Schreier.2014}, a top-down approach, with some components of Grounded Theory~\cite{Glaser.1967, Charmaz.2006, Strauss.2008} which operates in an inductive, bottom-up manner~\cite{Saldana.2014}. We used this two-pronged approach to systematically identify patterns and themes in our interviews. On the one hand, we aimed at superficially known but understudied aspects of hosting activities as reflected in our interview-guide questions and corresponding codes. On the other hand, the bottom-up component allowed us to identify phenomena which have not been noticed by previous research.

\subsection{Sampling Outline}
Our sampling strategy followed the general principle of maximal structural variation~\cite{Kleining.1982}, with the aim of capturing diverse behavior across providers offering different combinations of services. The goal was to expand the surface of investigation with respect to our main research focus: interventions following vulnerability notifications.
We approached this from two angles. First, we considered the potential for hosting provider intervention. Different types of hosting services imply varying degrees of provider control~\cite{tajalizadehkhoob2017herding}.
For instance, shared hosting typically grants customers fewer privileges while leaving a larger portion of the technical infrastructure under the HPO's control, compared to dedicated hosting. 
Second, we scoped our study to services known for suffering high levels of abuse, as these are more likely to receive vulnerability notifications. This supported our focus on the broader category of shared hosting services, which consistently show particularly high concentrations of abuse~\cite{tajalizadehkhoob2017herding, awpg-gps, tajalizadehkhoob2016apples}.
From this framing, we derived two key HPO characteristics likely to influence remediation behavior: the type of shared hosting service (e.g., web application, VPS) and the level of service management (managed vs. unmanaged). These two dimensions formed the basis of a sampling matrix, which we used to map the hosting landscape and identify specific providers for interview recruitment.

\subsubsection{Mapping the Landscape of Hosting Services}
We manually compiled a list of hosting provider organizations. Using corporate datasets~\cite{opencorp, crunchbase, dnb} was prohibitively expensive, and identifying organizations via IPs or Autonomous Systems from ranked website lists (e.g.,\cite{ruth2022toppling, le2018tranco}) was discarded due to bias toward global providers.
Instead, we identified companies by analyzing discussions on relevant Reddit communities (r/VPS, r/agency, r/webhosting, r/Hosting). Posts were selected if they discussed hosting services, excluded if tagged as advertisements or focused on out-of-scope topics. This process yielded 150 posts and a preliminary list of 197 companies.

Next, we manually reviewed the websites of these companies, discarding those offering only out-of-scope services or that no longer existed. Using a top-down content analysis of advertised services, we mapped offerings into the sampling matrix (service types and management). This map served then as the basis for our sampling process, as we anticipated that perspectives on VN remediation would vary across different segments. To capture this variation, we aimed to interview representatives from organizations operating in diverse areas of the hosting landscape.  
After this review, we finalized a list of 175 companies with a total of 716 distinct offerings (\new{available at}~\cite{data-collected}), spanning six high-level categories of hosting services. 
These are Dedicated Server, Virtual Private Server, Reseller Hosting, Shared Hosting, Website Builder, Web Agency.

\paragraph{\textbf{Hosting Services and VNs in Scope}}
\label{subsec:bg_scope}
In this work, we consider VNs reporting issues as web application code vulnerabilities~\cite{canali2013role, stock2016hey, stock2018didn}, misconfigurations~\cite{pletinckx2021out, roth2020complex, li2016you}, or internet services abuse, such as malware, phishing, or SEO infections~\cite{cetin2016understanding, vasek2012malware, hennig2022standing, stivala2024uncovering}. Consequently, our focus is on hosting services and providers operating closer to the application layer, such as VPS providers, shared hosting providers, and web agencies.

\subsubsection{Web Hosting Services}
\label{subsec:bg-hosting-services}
Hosting services in scope can be categorized in three high-level types, here ordered by required technical expertise.
 
\textit{Virtual Private Servers (VPS)} offer dedicated portions of shared server resources, ensuring better performance and scalability but requiring higher technical expertise for setup and maintenance.
\textit{Shared hosting} involves instead multiple users sharing a single server's resources, such as bandwidth, CPU, and memory, ensuring affordability and simplicity for website owners. Website setup is facilitated through graphical interfaces, making the service accessible even to users with low technical skills.

\textit{Web agencies} (or ``resellers'') act as intermediaries between end-users and hosting infrastructure. Their revenue comes from developing and managing websites, with customization being central to their business model. The term has a dual meaning: ``reseller hosting'' refers to a service offered by HPOs, while ``web agencies'' are the businesses using this service to deliver final products to customers.

Hosting services may be either managed or unmanaged. \textit{Managed hosting} services include provider assistance, for example with configuration or maintenance, as defined in service agreements, while \textit{unmanaged hosting} leaves these responsibilities to the client. 
The level of management is closely tied to the underlying hosting type and can also vary across providers, as per their terms of service. For a VPS, management may focus on low-level tasks, such as operating system updates. In contrast, managed shared hosting often extends to application-layer services, such as website setup and configuration, catering to less technically proficient users.

\subsubsection{Recruitment}
We used the service data to group companies by similar offerings. We represented each company’s services as a one-hot-encoded vector and applied the k-modes clustering algorithm to group them, using an empirically-chosen value of $k=13$ for an optimal balance of interpretability and granularity. Our clustering results mirrored the distribution of services in our dataset, where the majority (78\%) of companies provide more than one service in various combinations. We obtained five clusters of HPOs offering primarily one service (e.g., website builder, web agency), and eight clusters where HPOs either offered combinations of two or three services (e.g., shared hosting and VPS) or were primarily centered on one service with additional, less significant offerings.

We chose to start from companies offering a single service to reduce the complexity of HPO characteristics that could influence VN remediation. This decision was based on our categorization of service offerings and management. 
We recruited participants using a mix of cold-calling via available contacts (phone, email, ticketing portal, or chat, resulting in nine participants), snowballing and personal contact networks (still guided by our sampling map, resulting in nine participants), and professional social networks (as LinkedIn, resulting in six participants).
This resulted in a total of 24 hosting providers and web agencies from eleven countries. We describe hosting providers and individual participants in \Cref{sec:participant-info} and \Cref{tab:part-details}.

\subsection{Interviewing Procedure}
\label{subsec:interview-procedure}

Following established methodological recommendations for semi-structured interviews~\cite{lazar2017research, Witzel2012problem}, we developed an interview guide as a flexible framework to steer the data collection process.
The primary aim was to explore if and how hosting provider organizations handle vulnerability notifications. Thus, a large portion of the guide addressed VNs, investigating employee awareness, communication channels, and decision-making processes. 

Recognizing that VN handling may be shaped by broader factors, we included sections examining technical factors (e.g., infrastructure and tools), business factors (e.g., provided services and management), and organizational factors (e.g., internal procedures and management priorities)~\cite{li2019keepers, botta2007towards, dietrich2018investigating}.
To familiarize ourselves with industry procedures for security management, we relied on the NIST Computer Security Incident Handling Guide ~\cite{cichonski2012computer}, which helped shape questions on identified assets, perceived risks, and operational aspects, such as the use of playbooks (step-by-step procedures for employees) and the involvement of third-party entities in incident response.    
We report the complete interview guide in Appendix \ref{app:ig}. 

The interview guide was tested in two pilot interviews with hosting-acquainted colleagues at our research institution. The guide was iteratively refined throughout the study by incorporating insights from earlier interviews to partly re-focus on some newly emergent, relevant issues.
For instance, some questions were tailored based on whether the interviewee was a classical hosting provider or part of a web agency. When participants spontaneously introduced some of the topics in the course of the interview, we followed their lead allowing for a change in the order of discussed topics. 
However, the interview guide always contained a stable core of topics we discussed with all our participants throughout our entire study.
\new{While we initially posed broad questions to avoid biasing responses, the data ultimately stem from participant responses concerning application-level vulnerability notifications; participants who did not engage with these were considered unfamiliar with VNs, and the rare mentions of other notification types (e.g., OS or network) were excluded from the analysis as out of scope.}
This approach aligns with best practices in qualitative research, which emphasize adaptability to enrich data quality.

Nearly all interviews were conducted via Zoom\footnote{One interview was conducted in person due to the proximity of HPO's premises to the location of our research institution.} between Aug `24 and June `25, lasting between 60 and 90 minutes. 
Before the interview, all participants completed a questionnaire providing consent, demographic details, and information about their HPO, reported in \Cref{tab:service-offerings}.
All interviews were audio-recorded and transcribed using our in-house AI-based transcription tool. 
Most interviews were conducted in English, with some in Italian later translated for data processing (four via DeepL and three via our in-house AI-based transcription tool). All interview transcripts were anonymized by removing personal and HPO-specific information.

\subsection{Data Analysis}
The analysis of interview data combined a top-down approach of Qualitative Content Analysis~\cite{Mayring.2000, Schreier.2012, Schreier.2014} with elements of Grouded Theory's bottom-up methodology~\cite{Glaser.1967, Charmaz.2006, Strauss.2008}. An initial set of codes was derived from the topical areas and specific questions in our interview guide. It was clear from the very beginning that our interviewees talked about these issues, because we explicitly asked them about it. To avoid fully subsuming our rich data under predetermined categories, we also allowed for the creation of new codes in a bottom-up manner inspired by Grounded Theory's open-coding procedure. This approach enabled us to capture specificity and novelty emerging from the data while anchoring it within the broader topical areas represented by the initial top-down codes~\cite{Saldana.2014}.

Initially, three researchers (a computer scientist, a psychologist and a sociologist) independently coded four transcripts to identify discrete text units relevant to the research questions. To ensure consistency and resolve interpretive differences, we collaboratively reviewed and refined the initial codes through detailed discussions, fostering shared understanding and adapting codes as necessary. We did not calculate inter-coder agreement rate as our final agreement on codings approached 100\%~\cite{Klostermeyer2024Skipping, Klivan2024Everyone, huaman2024you, schmuser2024analyzing}.

Following this, the researchers conducted several iterations of descriptive coding, grouping open codes into broader thematic categories. This iterative process led to the development of increasingly abstract categories that captured key themes and patterns in the data, resulting in a preliminary codebook.
The codebook was then applied to five additional interviews, during which it was tested, refined, and expanded to include omitted codes, ensuring comprehensive data coverage. Finally, we used the finalized codebook (Appendix \ref{app:codebook}) to code the remaining fifteen interviews, with all coding performed using ATLAS.ti~\cite{atlasti}.

We conducted interviews until all HPO clusters identified via the mapping procedure were covered, and our analysis indicated saturation in identifying and describing phenomena relevant to the research questions. By the end of this process, no new themes or problems emerged, confirming that the studied phenomenon and its diversity were comprehensively described. 
The final code system was then used for a category-driven re-examination of the primary data. A comparative matrix was created to extract RQ-relevant information from each interview, enabling the identification of similarities and differences in how individual HPOs addressed the problems focused by our research questions.

\subsection{Ethical considerations}
During the recruitment phase, we ensured that each company was contacted only once. If no interest was expressed, the company was excluded from further follow-up. Additionally, we sent a maximum of two reminders to potential interviewees who showed interest.

We conducted the interviews via videotelephony, but only recorded audio tracks. These audio files were used exclusively for transcription and were deleted after the transcription was completed.
To protect participant confidentiality, we kept demographic data and personally identifiable information (e.g., details collected during recruitment) strictly separate from the study data. Personally identifiable information was deleted upon completion of the data-gathering phase. Anonymity was preserved through the assignment of unique participant ID numbers and the use of password-protected spreadsheets accessible only to the principal investigators. Additionally, transcripts were fully anonymized to remove references to individuals, company names, specific locations, associated organizations, and any identifiable tools or products developed by the company.

All participants provided informed consent prior to their involvement. As compensation, we offered each participant 50 USD. Further, the responsible ERB for our institution reviewed and approved our study procedure.

\begin{table*}
	\centering
	\footnotesize

    \begin{tabular}{l|l|c|l|l|l|l|l|c|l}
        \toprule
        HPO ID & \makecell{HPO\\size} & \makecell{CySec dept.\\ or person} & Primary education & Current role & Age & YoE & \makecell{Years\\in HPO} & \makecell{Multiple\\HPOs} & \makecell{Country} \\
        \midrule
        \HPOone{} & 11-50 & \CIRCLE & Computer Science & Front-end dev., Proj. Manager & 29 & 7 & 1-3 & $\star$ & IT \\ 
        \HPOfive{} & 51-200 & \CIRCLE & Computer Science & System Operator & 24 & 2.5 & 4-10 & ~ & DE \\ 
        \HPOten{} & 1-10 & \Square & Computer Science & CEO & 41 & 10 & 11-20 & ~ & IT \\ 
        \HPOeight{} & 1-10 & \CIRCLE & Industrial electronics & Responsible Executive & 59 & 24 & 11-20 & ~ & IT \\ 
        \HPOtwo{} & 1-10 & $\Diamond$ & Computer Science & Owner & 23 & 7 & 4-10 & ~ & NL \\ 
        \HPOfiftn{} & 201-1000 & \UParrow & Media and Film & Product Owner & 37 & 7 & 4-10 & ~ & BG \\ 
        \HPOseventn{} & 11-50 & \UParrow & Computer Science & Customer Success Engineer & 32 & 4 & 4-10 & ~ & IN \\ 
        \HPOtwtytwo{} & 51-200 & \CIRCLE & Electronics and Telecomm. & Technical support executive & 25 & 3 & 1-3 & ~ & IN \\ 
        \HPOtwtyseven{} & $> 1000$ & \UParrow{} & Computer Animation & Digital Fraud Analyst & 42 & 8 & 11-20 & ~ & USA \\ 
        \HPOtwtyeight{} & 11-50 & \UParrow{} & Computer Science & CTO & 42 & 11 & $> 20$ & ~ & BG \\ 
        \HPOtwtyfour{} & 201-1000 & \UParrow & History & CEO & 37 & 20 & 1-3 & ~ & UK \\ 
        \HPOthree{} & 1-10 & \Square & Telecommunications & CEO & 45 & 27 & $> 20$ & ~ & DE \\ 
        \HPOsix{} & 11-50 & \Square & Computer Science & Developer & 29 & 10 & 4-10 & ~ & DE \\ 
        \HPOeightn{} & 51-200 & \UParrow & Computer Science & Technical Support & 27 & 7 & 1-3 & 3 & PH \\ 
        \HPOtwtysix{} & 1-10 & \Square{} & Computer Science & CEO \& CTO & 38 & 22 & $> 20$ & ~ & AT \\
        \HPOtwenty{} & $> 1000$ & \UParrow & Computer Science & Linux System Administrator & 26 & 5 & $< 1$ & 2 & IT \\
        \HPOthirtn{} & 11-50 & \UParrow & Computer Science & Cloud Administrator & 37 & 5 & 4-10 & ~ & IT \\ 
        \HPOnine{} & 1-10 & \Square & Did not attend university & Owner & 45 & 22 & $> 20$ & ~ & DE \\ 
        \HPOeleven{} & 11-50 & \UParrow & Business and economics & Operations and maintenance & ~ & 10 & 4-10 & ~ & IT \\ 
        \HPOfourtn{} & 51-200 & $\Diamond$ & Computer Science & System Administrator & 31 & 5 & 1-3 & ~ & IT \\ 
        \HPOtwtyone{} & 1-10 & \Square & Computer Science & Customer Success Manager & 19 & 2 & 1-3 & ~ & PL \\ 
        \HPOtwtyfive{} & 1-10 & - & High School Diploma & CEO & 19 & 2 & 1-3 & ~ & IN \\ 
        \HPOtwelve{} & $> 1000$ & \UParrow & Computer Science & Senior Network Engineer & 28 & 10 & $< 1$ & $\star$ & DE \\
        \HPOtwtythree{} & 1-10 & \CIRCLE & Communication & CEO & 42 & 20 & 4-10 & ~ & DK \\
        \bottomrule
    \end{tabular}
    \caption{Participant details collected via survey. Legend for \textit{HPO ID}: [Type]-[Mgmt]-[ID], where: Type: WA = Web Agency, SH = Shared (Web Application) Hosting, VPS = VPS. Mgmt: M = Managed, U = Unmanaged, MU = Both. ID: incremental number. Legend for \textit{CySec dept.}: \CIRCLE{} participant;  $\Diamond$ another employee; \Square{} collective decision; \UParrow{} HPO security/abuse department; - no department/employee. Legend for \textit{Multiple HPOs}: $\star$ for previous employer only, or total number $N$ of employers discussed. Note: participant’s country may differ from the HPO's, e.g., for global enterprises.}
	\label{tab:part-details}
\end{table*}

\begin{table}
    \footnotesize
    \centering
    \setlength{\tabcolsep}{1.4pt}
    \begin{tabular}{p{1.5cm}|p{1.5cm}|p{1.5cm}|p{1.5cm}|p{1.5cm}|p{1.5cm}} 
    \toprule
        Managed services & VPS &  Reseller Hosting & Shared Hosting &  Web Agency \\
        \midrule
        Managed & {\HPOthirtn{}, \HPOtwelve{}, \HPOtwtythree{}, \HPOtwenty{}, \HPOtwtytwo{}, \HPOtwtyeight{}} & {\HPOfiftn{}, \HPOseventn{}, \HPOtwtytwo{}, \HPOtwtyseven{}, \HPOtwtyeight{}} & {\HPOtwo{}, \HPOfiftn{}, \HPOseventn{}, \HPOtwtytwo{}, \HPOtwtyseven{}, \HPOtwtyeight{}, \HPOtwtyfour{}, \HPOthirtn{}} & {\HPOone{}, \HPOfive{}, \HPOten{}} \\ \hline
        Unmanaged &  {\HPOnine{}, \HPOeleven{}, \HPOfourtn{}, \HPOtwtyone{}, \HPOtwtyfive{}, \HPOtwelve{}, \HPOtwtysix{}} & {\HPOsix{}, \HPOtwtysix{}, \HPOnine{}, \HPOeleven{}, \HPOtwtythree{}} & {\HPOthree{}, \HPOsix{}, \HPOeightn{}, \HPOfourtn{}} & \HPOeight{} \\ \hline
        \rowcolor{green!20}{Management outside contract} & \HPOtwelve{}, \HPOtwtythree{} & ~ & \HPOthree{}, \HPOsix{}, \HPOtwtysix{}, \HPOtwenty{} & ~  \\ 
    \bottomrule
    \end{tabular}
    	\caption{Distribution of HPOs' services by support levels (rows) and service types (cols). The row \textit{management outside of contract} is a new finding, absent from initial market mapping.}
	\label{tab:service-offerings}
\end{table}

\subsection{Participant Data and HPO Information}
\label{sec:participant-info}
Our interview study included 24 participants from eleven different countries, reported in \Cref{tab:part-details}. 
Participants represented a wide variety of positions and represented companies varying significantly in size, ranging from small (less than 10 employees, nine companies) to mid-sized (11–50, 51–200, 201–1,000, twelve companies) to large (over 1,000 employees, three companies). 
Participants demonstrated very diverse technical skillsets tied to their roles. In a nutshell, their tasks can be divided in three categories: leaders (ten participants), customer-facing (seven participants), and backend roles (seven participants).

Regarding the services offered, the study captured a wide range of market coverage, reported in \Cref{tab:service-offerings}. Web application hosting services (``shared hosting'') were the most common, provided by 12 companies. Eight companies offered VPS solutions, while four operated as web agencies. 
When examining the nature of hosting management, the split was almost even: eleven companies provided managed hosting, while ten operated on an unmanaged basis, and three offered both options.

While most participants focused on a single company during the interview, four offered insights into more than one. Two of them (\HPOone{}, \HPOtwelve{}, marked with $\star$ in \Cref{tab:part-details}) discussed previous employers, which were more relevant to our study than their current roles.
The remaining two reported experiences from multiple companies: \HPOeightn{} referenced three organizations, and \HPOtwenty{} discussed two. 
\new{None of the HPOs discussed by these participants overlapped with those described by others.}

We noted partial overlap between \HPOnine{} and \HPOsix{}, as \HPOsix{} is a spin-off from \HPOnine{}, established to diversify their product offerings. Moreover, we remark that the service categorizations reported in \Cref{tab:service-offerings} follow the companies' self-descriptions, though their actual offerings and management levels vary and are not always consistent across companies.
Lastly, \HPOfourtn{} is a non-profit organization in the education sector, whose services align closely with those of typical hosting provider organizations, while \HPOeightn{} and \HPOtwenty{} both provide registrar services alongside of hosting.

\subsection{Structure of Findings}
The following sections \ref{sec:vulnerability-notification} through \ref{sec:tech-infrastructure}  address our research questions with case-specific examples and conclude with key takeaways. \Cref{sec:implications} discusses main findings, offers actionable insights for future notifications, and suggests directions for further research.
We use \textquote{angular parentheses} to report verbatim quotes from interviews.
\section{Notification Channel and Message}
\label{sec:vulnerability-notification}

This section presents our findings on the unprompted communications received by HPOs from external actors regarding the security and privacy of their infrastructure or hosted customer instances, addressing the first part of \textbf{RQ1} on \textit{How are vulnerability notifications received}. Our questions focused on factors explored in prior work, such as reachability and email characteristics, while allowing participants to share their experiences with all forms of unprompted communication.

\subsection{Receiving VNs}

\subsubsection{Awareness of VNs}
All but three participants (\HPOone{}, \HPOthirtn{}, \HPOseventn{}) were familiar with the concept of vulnerability notification.
Most participants regularly engaged with them (e.g., \HPOtwo{}, \HPOthree{}, \HPOfive{}, \HPOnine{}, \HPOtwelve{}, \HPOtwenty{}) and had a positive or neutral view of this mean of communication. 
Conversely, a subset of participants expressed skepticism regarding the intent of VN senders (\HPOten{}), or doubted that external actors could identify vulnerabilities they were not already aware of (\HPOfourtn{}, \HPOfiftn{}). 
For others, the idea of receiving notifications from \textit{unrelated} third parties was unfamiliar, expecting to receive vulnerability reports only from established business relationships or trusted communication channels (\HPOthirtn{}, \HPOseventn{}, \HPOtwtyfive{}).

\subsubsection{Reachability}
Participants were asked about the channels available for external actors to report VNs. Responses highlighted seven distinct communication endpoints and six different channels, with some participants identifying multiple methods.

WHOIS was the most frequently mentioned channel (seven HPOs), primarily among organizations managing their own infrastructure. 
One large company reported primarily receiving machine-readable abuse reports as a paid service, and two more reported preferring submissions via the company portal. Interestingly, one VPS provider reported only receiving abuse notifications through their provider, as they rent all infrastructure and don't own any IPs.

Web agencies provided contact details either in website credits or on a separate page, such as an ``Impressum'' or a \texttt{security.txt} file (one agency). None relied on WHOIS for communication.
Small web agencies focusing on website development and management relied heavily on their hosting providers to inform them of issues, often due to a lack of in-house expertise or lack of alignment with their business model and setup (\HPOone{}, \HPOten{}).

\begin{figure}
  \centering
  \includegraphics[width=\columnwidth]{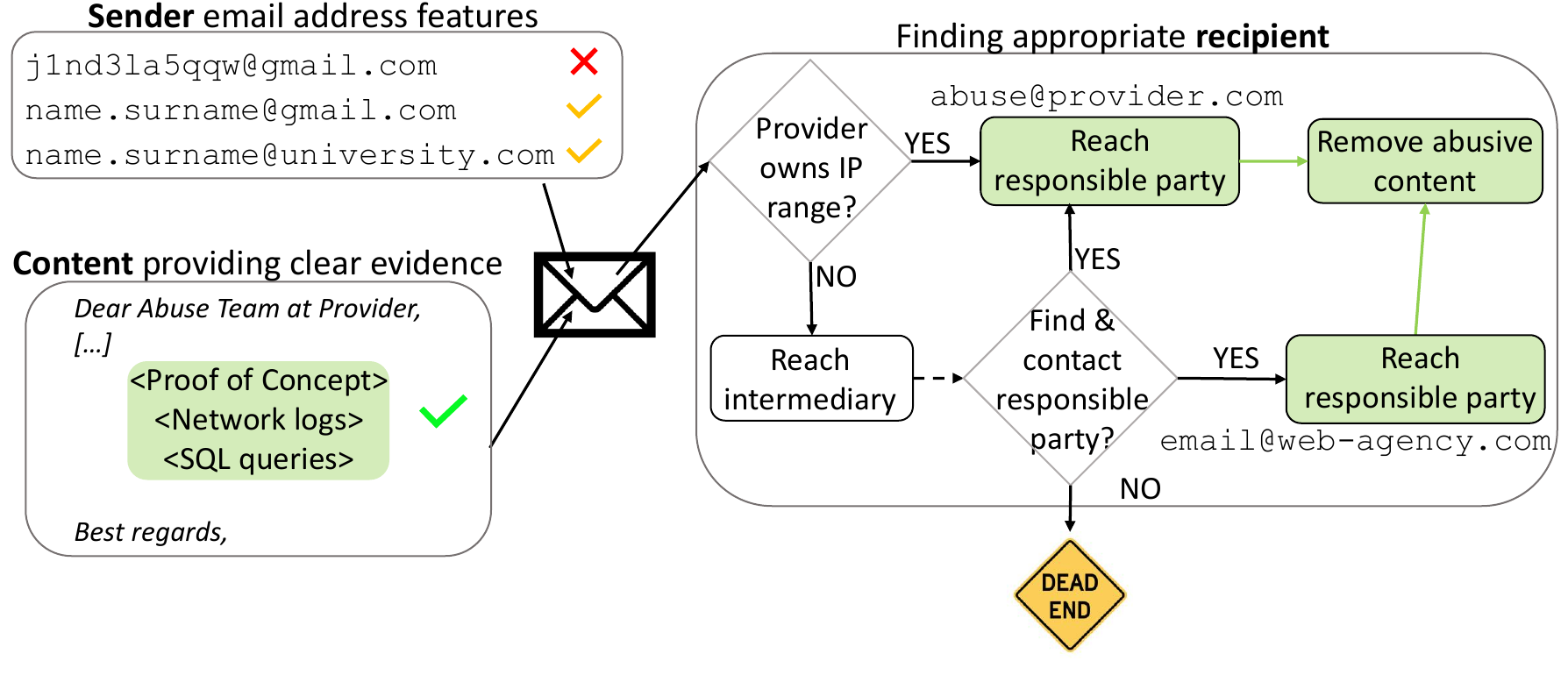}
  \caption{
     VN creation process based on provider feedback. The sender’s email is only relevant if it raises red flags; the body’s evidence is most important. On the left, the diagram shows how the email’s routing depends on the recipient: if the HPO owns the IP range (per WHOIS), they can be contacted directly. Otherwise, the VN is sent to the infrastructure provider (i.e., the IP range owner, e.g., a data center or cloud provider), who acts as an intermediary and must forward the message to the affected HPO.
  }
  \label{fig:vn-workflow}
\end{figure}

\subsection{Content and Sender Characteristics}
\subsubsection{Senders of VNs}
HPOs receive security-related communications from a diverse range of senders. Eight participants reported receiving notifications regularly from various entities, which can be grouped into three main categories.

First, four participants (\HPOsix{}, \HPOnine{}, \HPOtwelve{}, \HPOeightn{}) mentioned receiving security advisories or investigation requests from government agencies, CERTs, or police departments.
Then, four participants mentioned established commercial companies focused on spam, malware, phishing (e.g., Netcraft), or legal issues as brand protection and DMCA reports. These senders were seen as reliable and their notifications were generally addressed promptly due to their established reputations.

Three participants reported receiving reports from private individuals, referred to as \textquote{white hat hackers} (\HPOfive{}) or \textquote{technician[s]} (derogatory, \HPOten{}). However, opinions on these reports were divided. While some participants appreciated these notifications, others found it challenging to distinguish genuine reports from unsubstantiated claims or requests for monetary compensation. One participant summarized the skepticism:
\textquote{[``Security researcher''] is not an official protected term, a lot of people call themselves security researcher, which might be the 14 year old from Bangalore who thinks revealing the web service version number is dangerous} (\HPOnine{}).
Several participants, including both small and large HPOs, reported experiencing high volumes of spam in their VN inboxes. The cause remained unclear, as both affected and unaffected participants used spam filters and public \texttt{abuse@} addresses. Abuse reports from private individuals were uncommon: only three participants recalled interactions with academics, while others mentioned cases involving individuals directly affected by abuse on their platforms, such as victims of DDoS attacks or impersonation.

\subsubsection{Email Content and Metadata}
Among participants familiar with VNs, the criteria for evaluating reports were a key discussion point. While prior work highlighted email features (style, tone, metadata), most participants downplayed their importance. Only a few preferred formal email addresses; most based decisions on technical content, prioritizing reports with evidence (e.g., logs or Proof-of-Concept), and directly examined of the reported website.
\HPOten{}, \HPOfive{}, and \HPOtwtythree{} report receiving emails from private addresses, raising suspicion: \textquote{[T]his is a cryptic email address [...]. So everything looks really, really phishy. And afterwards you say, okay, thanks for your report} (\HPOfive{}). 
\HPOtwtythree{} also addressed the case of random email addresses, which prompted extra scrutiny: \textquote{Some people [...] send it from some random streaming of characters @gmail.com. In which case, we take a look at it a little bit closer}.
In summary, most providers accept personal email addresses, but emails from random-looking senders are examined more critically, with the final assessment based on the report’s content.

\begin{tcolorbox}[colframe=black, colback=white, boxrule=0.3pt, arc=0mm]
\paragraph{\textbf{Takeaways}} Our findings do not indicate significant roadblocks concerning awareness and posture towards VNs. While we observed that reachability can be an issue, given by hosting setups involving middle layers as registrars, resellers, and outsourced services, most providers still indicated WHOIS as the right source for their contacts.
\Cref{fig:vn-workflow} illustrates the creation of a VN based on the feedback from our providers.
Providers reported being contacted regularly by security companies, which creates established communication channels and patterns. Rarely receiving notifications from individuals, such as researchers, might impact HPOs' perceptions of these communications, potentially making them more suspicious. However, participants noted that reports including strong evidence of issues were always considered, regardless of sender reputation and email metadata.
\end{tcolorbox}

\section{VN Handling Procedures}
\label{sec:organizational-structure}

This section examines the internal processes triggered when a security issue is detected or reported to an HPO, addressing the second part of \textbf{RQ1} on \textit{How are vulnerability notifications processed}. Participants described how security event management was handled in their organizations, including the roles and responsibilities of security teams, remediation procedures, internal communication protocols, and interactions with customers or external stakeholders.

\subsection{Internal Handling Procedures}
HPOs can adopt structured methods to ensure consistent handling of common problems, sometimes referred to as ``playbooks'', though the level of formalization varies significantly. 
We investigated the presence of formalized procedures regulating vulnerability notification handling.

We observed that most providers do not follow a strict or formalized process for handling vulnerability notifications, though some have more structured approaches than others.
For example, operators at \HPOtwelve{} and \HPOtwtyseven{} create playbooks for managing security events as part of their regular tasks, while \HPOtwenty{} reported having a playbook to address VN reports at their previous employer.
Other HPOs standardized interventions in customer spaces via internal \textquote{wikis} (e.g., \HPOthree{}, \HPOfiftn{}, \HPOseventn{}, \HPOeightn{}). 
 
Most participants described having minimal or no formal guidelines for handling security events. For example, at \HPOfive{}, security decisions are made based on log interpretations, while at \HPOfiftn{} and \HPOtwenty{}, operators are encouraged to address issues leveraging their own skills.
\HPOnine{}, also owner of a spin-off, \HPOsix{}, relies on a habitualized~\cite{Bourdieu1977Outline} cybersecurity awareness grounded in his employees' professional socialization within tech-savvy hacktivist backgrounds or open-source programmer communities. Thus, they are assumed to be highly sensitive to S\&P topics: \textquote{The majority of my team members are coming from like hacker spaces and generally the open source community. And they already have a deeply ingrained approach, they discover security problems in open source software on a daily basis and handle them in appropriate ways. And I think they do this in the best and most responsible way}.

\subsubsection{Impact of Internal Organizational Structures}
We interviewed IT professionals across a wide range of roles, including executives, abuse analysts, and support staff, from both small and large hosting providers, ensuring a diverse and balanced perspective across organizational scales.
This diversity allowed us to examine who handles VNs, the steps involved, and the interactions among professionals within hosting organizations. This was particularly evident in larger companies, where specialization and compartmentalization are more common. In such cases, VNs are typically managed by a dedicated abuse department with access to the \texttt{abuse@} mailbox. These operators evaluate VNs in a self-contained process that rarely involves other departments, and may also develop internal playbooks for handling specific types of application-level abuse.
When VNs involve legal concerns, as fraud or DMCA complaints, they may be forwarded to the legal department. In rare instances, when VNs relate to infrastructural issues, such as those at the IP level, the abuse team may collaborate with infrastructure-focused employees, such as DevOps or Site Reliability Engineers.

Customer support also plays a role in abuse handling as the first point of contact for clients. While most interactions involve routine setup or configuration, support agents may occasionally assist clients affected by exploits. In complex cases, and where company structure permits, support staff can escalate issues to more experienced personnel, such as ``Level 2 support'' or the abuse team. However, in practice, the two departments usually operate independently.
Participants reported no interdepartmental friction or procedural obstacles in addressing VNs. In smaller companies, roles are often consolidated, with a single employee managing the entire process.

\subsubsection{Procedures Established by Certifications}
Certifications are formal attestations granted by recognized standardization bodies, indicating that an organization’s processes or products meet specific industry requirements. When a company certifies its procedures or software, it means these have been evaluated and shown to comply with the defined criteria of the relevant standard (e.g., ISO/IEC 27001 for information security management). Certification can apply to all internal procedures or only selected ones, depending on organizational priorities. In this context, we sought to determine whether holding certifications related to security or incident response had any impact on how HPOs manage VN handling.
Few participants (\HPOtwelve{}, \HPOthirtn{}, \HPOtwenty{}) reported getting one or more certifications to access specific market segments, such as public administration (\HPOtwelve{}, \HPOtwenty{}) or Limited Liability Companies (\HPOthirtn{}).
\textquote{We are about to take the ISO 27001 certification, [...] Because, you know, it is fundamental for this type of thing. [...] Because, above all, customers ask for it} (\HPOthirtn{}).
Certifications were however not mentioned as a factor playing in procedures for remediating issues or vulnerabilities in customer spaces, situations for which participants reported following other legal frameworks (especially nation state's or GDPR). These participants reported certifying procedures such as \textquote{datacenter encryption} (\HPOtwelve{}) or \textquote{SLA requirements and disaster recovery} (\HPOthirtn{}). Notably, \HPOeight{}, a web agency developing software for the public administration, reported having no constraint on the application software.

\subsection{Procedures Involving External Stakeholders}
We examined potential remediation challenges arising from a multi-stakeholder setup. Participants did not report dependencies on external entities involved in remediation, such as CERTs or Managed Security Service Providers. 

\subsubsection{Interactions between Hosting Providers, Web Agencies, and Final Customers} One notable multi-stakeholder dynamic is the ``reseller hosting'' model (introduced in \Cref{subsec:bg-hosting-services}). We investigated the responsibility and authorizational boundaries between hosting providers and their clients (``resellers'', or web agencies). \HPOfiftn{} and \HPOseventn{}, both managed hosting providers, described their admin panel as having a fine-grained access control system that enables both intermediaries (web agencies) and end-users (web agency clients) to act within customer spaces, effectively removing authorization barriers. While this paints a positive picture, none of the web agencies outsourcing hosting (\HPOone{}, \HPOeight{}, \HPOten{}) reported using managed shared hosting. Instead, they preferred service packages without customer support, shifting the burden of remediation to themselves.
A few participants described a strict authorization boundary in both directions (e.g., \HPOtwo{}, \HPOtwtyeight{}, \HPOfive{}, \HPOeight{}). Hosting providers noted that some resellers enforce clear separation of responsibilities (\textquote{We do not interfere there, and actually many of them do not allow us to touch [their] products}, \HPOtwtyeight{}). Conversely, web agencies reported losing contact with clients after the initial website development, as ongoing management was transferred to the client’s internal IT department.

Finally, web agencies (e.g., \HPOfive{}, \HPOten{}) might experience issues when onboarding of customers with externally developed websites. In such cases, \HPOten{} notes that their typical solution is to freeze the codebase until the customer agrees to pay for a complete rewrite, even if the website runs outdated or vulnerable code.

\subsubsection{Procedures at Registrars}
Domain registrars are companies authorized to manage the reservation of internet domain names and typically operate separately from hosting providers. In cases of abuse, registrars may receive complaints related to domain ownership or misuse, where they can react by, e.g., disabling domain resolution by web clients. However, enforcement actions like content removal fall under the responsibility of hosting providers.

Two participants, \HPOeightn{} and \HPOtwenty{}, also acted as domain registrars, providing unique insights absent from accounts by other interviewees.  
We report their valuable insights, acknowledging the limited generalizability.
The participants reported checking all incoming emails, regardless of sender or email features, considering as only criterion whether the domain or content was under their control. Providers reported taking immediate action in case the domain hosted illegal material (e.g., phishing, copyright violations, child pornography, or other breaches of local law), suspending the domain and notifying both customer and hosting provider. For all other types of issues, HPO behaviors differed depending on their business model: \HPOeightn{}, offering unmanaged services, left the website untouched, even if obviously compromised. \HPOtwenty{}, offering both managed and unmanaged services, evaluated whether the reported flaw posed risks to infrastructure stability, other customers’ instances, or sensitive data security. If none of these applied and the customer was not a premium client, they sent an email notification to the client without further action.

\begin{tcolorbox}[colframe=black, colback=white, boxrule=0.3pt, arc=0mm]
\paragraph{\textbf{Takeaways}} Few HPOs reported having a clearly defined process outlining specific steps and criteria for handling vulnerability notifications, while most of them follow semi-formalized procedures which participants generally describe as flexible and not limiting to their operations. Handling of incoming reports is typically left to the discretion of individual operators, where most rely on unwritten but commonly followed rules. Internal HPO structure and division of responsibilities was not reported as a roadblock by participants working in medium to large HPOs.
Few hosting providers reported having clear authorization boundaries when dealing with web agencies as intermediaries, in which cases they cannot take action directly and simply forward the received VN. Most web agencies confirmed that they rely on hosting providers to be informed about website issues.
\end{tcolorbox}

\section{Deciding Whether to Intervene}
\label{sec:criteria-for-decision}
After establishing that some VNs are received and that handling procedures are often informal, we asked participants what actions they would take and why, addressing \textbf{RQ2}. Their decision-making criteria largely fell into three categories, outlined below. Participants also reported challenges in determining whether to intervene, as well as exceptional circumstances that prompt remediation.

\subsection{Type of Vulnerability or Issue}
We investigated whether the type of reported vulnerability influenced providers' remediation decisions. In general, action was taken when there was evidence of ongoing malicious activity, such as phishing, spam, malware delivery, or distribution of illegal content. Almost all participants reported they would respond by removing the malicious content, and potentially blocking parts of the website or taking the website offline. Additionally, they notified affected customers, using methods ranging from simple alerts to more engaged outreach prompting customer-side remediation.

Alerts concerning \textit{potential} exploits, such as vulnerabilities in web applications, were treated differently. VPS providers, whether managed or unmanaged, stated they were not contractually obligated to oversee application software installed by customers, even when deployed using tools provided by the HPO. As a result, they did not address security and privacy issues at the application layer. This approach was described by \HPOsix{} as \textquote{the infrastructure-customer divide}, defining the boundary of responsibility between provider and customer. These providers focused on maintaining infrastructure stability by managing elements such as operating systems, databases, and programming environments, and responded to reports involving Denial of Service, port scanning, or email spam, but drew a firm line at application-level issues.

In the case of shared hosting, unmanaged providers described a hands-off approach to all web application concerns, stating \textquote{it's basically out of scope for us to take care of that [...]. Then you've got a shell there. Have fun.} (\HPOsix{}).
Managed service providers expressed a similar stance: \textquote{You get a place where you do not think of operating system upgrades, control panel upgrades, compatibility issues, [...] everything is covered. But it's that's where we our job ends and it's your responsibility to upload the software and after that to update it and keep it to date and safe} (\HPOtwtyeight{}). This underscores a strict division of responsibility between hosting infrastructure and code management, further echoed by another provider: \textquote{[But if] the issue with the application [is] like, in the coding, so we do not provide any development-related support} (\HPOtwtytwo{)}.

Operators reported two reasons behind this approach: first, the low cost of their offerings, which does not compensate for the time spent remediating and, additionally, the \textquote{bajillion} tickets they need to deal with daily (\HPOtwtyseven{}), requiring them to minimize the time spent for each of them.

\subsection{Legal Constraints}
Legal obligations play a significant role in motivating hosting providers to take action. 
For example, there are cases where the operator must apply a patch or update to avoid legal liability, or instead, they refrain from intervening, even when capable, because taking action could expose them to legal liability.
Two examples illustrate these dynamics. 
Participants generally reported no awareness of legal requirements concerning the maintenance of hosting infrastructure or software. \HPOtwenty{}, for instance, confirmed the absence of such regulations but emphasized that existing laws do penalize service providers in the event of a data breach. To reduce this risk, the company has adopted an internal update policy aimed at preventing incidents that could lead to legal consequences: \textquote{[T]here is no law that tells you that all systems must be in the last stable version, but clearly it tells you that if, because of the system, [because of] a vulnerability... personal data, sensitive data, and so on, get stolen, you are responsible} (\HPOtwenty{}).

The second example concerns unmanaged contracts. In these cases, the legal obligation typically extends only to notifying the client about the issue (e.g., \HPOthree{}, \HPOfive{}, \HPOeightn{}, \HPOtwtyeight{}). Once the report is delivered, responsibility shifts to the client, and the operator is no longer involved.
Applying code patches or modifying incompatible modules is not only unnecessarily costly when the issue does not affect others, but also risky, especially for highly customized setups. An unfamiliar operator might unintentionally break the site: \textquote{[I]n the [BUSINESS UNIT] where I am, there are about 50 people more or less, and the clients are.. hundreds of thousands... so... Understanding and taking action on every single personalization that can be done on the various websites, CMS websites [...] would be madness.} (\HPOtwenty{}). As a result, proactive remediation is not an option, since making changes would make the operator legally responsible. In managed contracts, when customers request a fix, some operators may choose to intervene but require the customer to sign a waiver before making any changes to the hosting environment.

\subsection{Customer-Specific Factors and Interactions}
\subsubsection{Interactions with Customers}

In some cases, customers react to vulnerability notifications and reach back to their hosting providers.
Participants expressed mixed feelings about customers raising such issues (e.g., \HPOthree{}, \HPOfive{}, \HPOten{}). They frequently reported that customers lacked the necessary understanding, demanded significant time for explanations, and created an unnecessary burden on their operations. Customer service is particularly costly for providers, creating challenges for both unmanaged and managed hosting services. Many HPOs face risks of financial losses due to the high cost of support agents relative to their low prices.

Many HPOs (e.g., \HPOthree{}, \HPOfive{}, \HPOeleven{}, \HPOtwelve{}) report that customers are often unwilling to pay security services (e.g., website cleanup). 
To mitigate this, providers like \HPOsix{} and \HPOnine{} focus heavily on creating detailed wiki articles to assist customers, without direct support, while \HPOfiftn{} requires users to consult their knowledge base before contacting support:
\textquote{[I]f everybody comes to tickets, our tech support team will be simply overwhelmed} (\HPOfiftn{}).

\subsubsection{Management Defined by Contract}
Companies offering unmanaged services limit customer support actions to written suggestions, such as troubleshooting steps or links to online resources (e.g., \HPOsix{}, \HPOeightn{}). A few of them reported expectations mismatches with customers on the level of management happen, usually solved via polite explanations (\HPOsix{}, \HPOnine{}) or by proposing additional paid services (e.g., \HPOfive{}, \HPOtwenty{}, \HPOtwtyfour{}), with some, like \HPOeightn{}, upselling services with sales-based bonuses. 
Managed service providers offered broader support, though intervention typically required a customer request and varied broadly by provider. 
Support agents often demonstrated goodwill in helping customers by updating software (e.g., \HPOone{}, \HPOtwo{}), restoring backups, or removing affected content upon request  (e.g., \HPOthirtn{}, \HPOfiftn{}, \HPOseventn{}). 
No provider mentioned actively inspecting for vulnerabilities, and only one (\HPOone{}) reported reviewing plugin code to ensure functionality. 

Code-level remediation was rare and generally offered when contractual services included custom web application development (\HPOtwo{}, \HPOten{}, \HPOtwelve{}, \HPOfive{}). For example, \HPOten{} described a complex, manual remediation of a hacked WordPress site redirecting visitors to malicious content. This intervention followed a customer complaint triggered by a Google SafeBrowsing warning. However, the same provider admitted to ignoring outdated PHP versions with known vulnerabilities and websites lacking HTTPS. This suggests that customer involvement may have strongly influenced their decision to act.

Conversely, business concerns might motivate providers not to address security issues. For instance, \HPOfive{} and \HPOten{} develop custom software, generating revenue by offering improvements as new product features. \HPOten{} notes that while their custom CMS is regularly updated and patched, updates are withheld from customers due to their unique customizations and to encourage them to purchase updated installations.

\subsubsection{Exceptional Proactive Remediation}
Occasionally, hosting providers take proactive measures to address compromised customer instances without waiting for them to reach back. These actions, such as disabling compromised websites or sections of them, are taken to safeguard the overall infrastructure's security and ensure fair resource allocation among all customers. This approach is often necessary because customers are slow to respond to notifications (e.g., \HPOthree{}, \HPOsix{}, \HPOnine{}, \HPOtwenty{}).

The motivations for such proactive efforts are often rooted in business strategy. For example, \HPOthree{} emphasizes high-quality customer service as a key differentiator but seeks to reduce time-intensive customer calls by addressing issues proactively. \HPOtwelve{}, which blends managed and unmanaged services, may exceed their contractual obligations to maintain goodwill and strengthen customer relationships--a decision driven by their administrative department. \HPOsix{}, \HPOnine{}, and \HPOtwtythree{} on the other hand, may proactively address compromised customer instances out of a commitment to prevent internet abuse, a stance shaped by their organizational and ethical principles: \textquote{We consider this as part of our responsibility that we not only have for our customers, but also for the rest of the internet. We don't want our machines to attack other machines} (\HPOnine{}).

\begin{figure}
  \centering
  \includegraphics[width=\columnwidth]{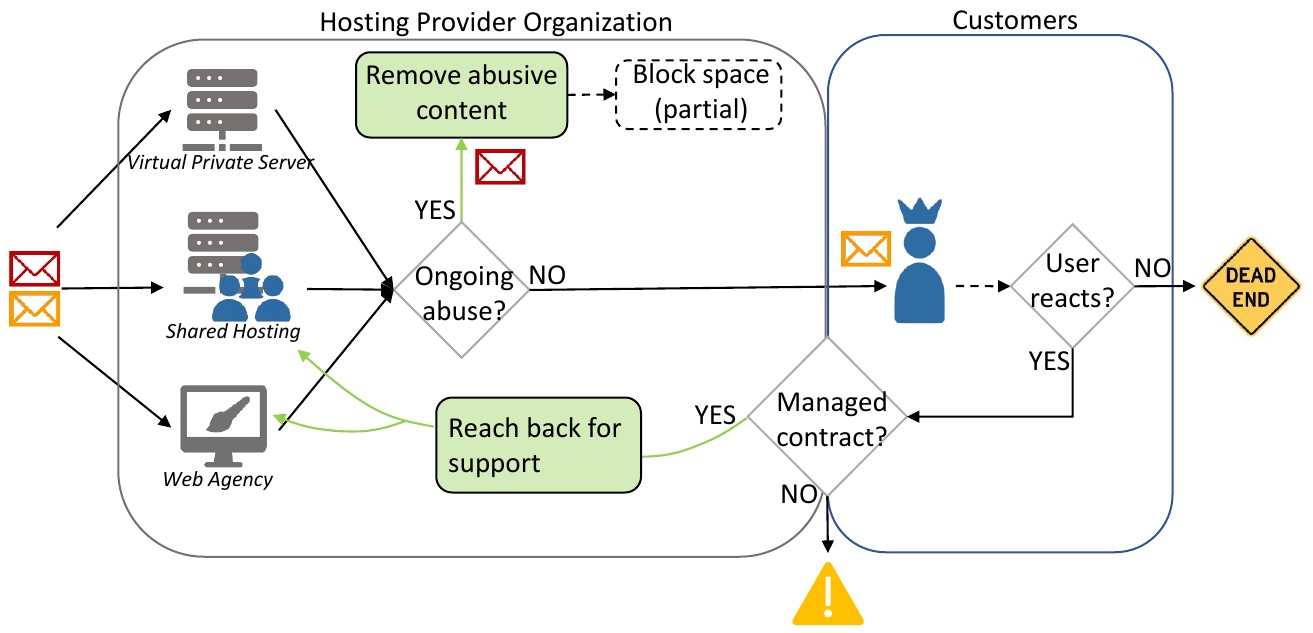}
  \caption{
    Notification handling workflow. Incoming reports are analyzed by abuse teams and abusive content taken down, with further actions depending on the HPO. Reports are otherwise forwarded to customers, and escalation to customer support depend on user action and contract terms. 
  }
  \label{fig:abuse-handling}
\end{figure}

\begin{tcolorbox}[colframe=black, colback=white, boxrule=0.3pt, arc=0mm]
\paragraph{\textbf{Takeaways}} \Cref{fig:abuse-handling} illustrates the VN handling workflow that emerged from our interviews. Most providers indicated their abuse team promptly addresses abuse such as phishing, malware delivery, or distribution of illegal content, regardless of the type of hosting and of the degree of service management. Reports concerning other types of web-application issues are forwarded to customers. In fact, our findings reveal a common understanding of service boundaries: customers, or web agencies in their stance, are responsible for code maintenance, while hosting providers focus on infrastructure and service availability. 
Moreover, the type of service provided affects the level of customer support, with some providers supporting users in remediations to a larger extent. 
Finally, providers are also influenced by factors that extend beyond the content of the VN, as the company’s ethical position on addressing internet abuse, the perceived value of long-term customer satisfaction, and the motivations of individual support agents.
\end{tcolorbox}
\section{The Role of VNs in Infrastructure Security}
\label{sec:tech-infrastructure}

While most hosting providers do receive vulnerability notifications, our earlier findings indicate that these are often not prioritized in practice, highlighting a misalignment between provider security goals and those of researchers. To better understand this disconnect, we examined providers’ awareness of abuse within their infrastructure, the security measures they employ at a technical level, and their perceptions of risk related to compromised customer instances.

\subsection{Provider Awareness of Abuse and Security}
We began by examining the extent to which hosting providers are aware of abuse and of the security posture of their customers operating on their infrastructure. Our goal was to assess whether visibility given by VNs could offer providers insights they currently lack, demonstrating practical value in enhancing their situational awareness.

Most hosting providers focus on maintaining visibility and control over their infrastructure to ensure reliable service and uphold service-level agreements. Providers who own their hardware (e.g., \HPOtwtyeight{}, \HPOtwenty{}, \HPOfive{}, \HPOfourtn{}, \HPOtwtythree{}) typically implement routines that monitor network activity, I/O, RAM, and other resource usage. For some, these monitoring systems also serve as early warnings for potential security incidents, but it usually falls to a skilled operator to determine if unusual activity reflects legitimate use or malicious behavior (\HPOfive{}).
HPOs that do not manage their own hardware  (e.g., \HPOone{}, \HPOeight{}, \HPOten{}, \HPOtwtyfive{}) often depend on tools supplied by their upstream hosting partners, using these to gain some operational insight. In some cases, structured monitoring is minimal or absent, with providers relying mainly on customer reports to identify problems (\HPOten{}, \HPOtwtyone{}).

Overall, the primary goal of monitoring is to maintain infrastructure performance, not to detect abuse or assess the security of customer spaces. As a result, if customer instances are compromised but do not disrupt service or exceed expected resource usage, these incidents frequently go unnoticed and unaddressed. Only a minority of providers report actively assessing the security posture of their customers’ environments (\HPOthree{}, \HPOfiftn{}).

\subsubsection{Securing Hosting Infrastructure}
Most providers secure their infrastructure’s by employing protections such as DDoS mitigation, firewalls, and isolated virtual networks (\HPOeleven{}, \HPOthirtn{}). Many use Linux permissions and least privilege principles to prevent customer infections from escalating  (\HPOthree{}, \HPOsix{}, \HPOnine{}), though implementation varies widely based on provider size and whether hosting is in-house or outsourced. Only few providers monitor for abuse higher up the stack, with some reducing risk by by preventing web app code edits: \textquote{the WordPress code itself is locked. Like no one can edit it. [...] So that's something that is totally secure and locked down} (\HPOseventn{}). Alternatively, hosting providers use web application firewalls, or run signature-based file scans (\HPOtwo{}, \HPOthree{}). Web agencies may also deploy code-level protections (\HPOeight{}, \HPOten{}).

\subsection{Perception of Risk from Customer Spaces}
HPOs described a range of technical measures used to secure their infrastructure, with a primary focus on ensuring resilience and service continuity. Because their responses centered on infrastructure security, we examined their perceptions of risks associated with customer spaces. Our goal was to understand the reasons behind their limited engagement, including whether this results from a perceived low risk or deliberate trade-offs.

\subsubsection{Perceived Risk from Customer Instances}
\label{subsec:perceived-risk}
Customer instance compromise is common, \textquote{we see WordPress exploits the whole time}, as \HPOtwtyfour{} reports. However, providers expressed confidence in their infrastructure’s ability to contain such incidents, keeping systems stable and functional. As \HPOnine{} noted, \textquote{our infrastructure does not care about phishing}.
Only one participant, \HPOone{}, recounted a full compromise of their hosting space after they transitioned to a less managed setup to meet the demands of a larger client.

VPS instances are typically compromised due to customer misconfigurations or are acquired directly for illicit use (e.g., \HPOtwelve{}, \HPOeleven{}, \HPOtwtythree{}). In shared hosting, abuse often stems from  outdated websites and plugins, malicious uploads, cracked plugins, or weak or leaked admin passwords. 
Although attacks on popular open-source web applications are well known, providers continue to support them and deal with the resulting abuse. For example, \HPOten{} accommodates customers who request WordPress sites, despite acknowledging they are frequent targets: \textquote{[E]very now and then we get that client who insists on having WordPress no matter what, so, to avoid losing the client or the project during the quote phase...}.

Web agencies expressed a different perception of risk compared to shared hosting and VPS providers.
For them, the main concern is protecting data, as they see little incentive for attackers to target custom code given the high cost and low potential gain.
\HPOten{} believes that custom-built web applications are inherently secure due to their unique nature, leveraging a form of ``security by obscurity''. \HPOfive{} similarly claims they have never had a hacked website, although they experienced a sophisticated data exfiltration attack. 
While custom code can reduce exposure to automated attacks, \HPOten{} and \HPOfive{} also acknowledged that it introduces risks from implementation flaws and overlooked interactions. Nonetheless, they believe the resources required to test for vulnerabilities across many customized deployments are not economically justifiable.

Finally, risk perception also depends on what providers consider worth protecting. \HPOten{} for instance, disregards HTTPS implementation on sites they believe contain nothing valuable. Similarly, \HPOeleven{} suggests that security should reflect business value, stating that loose measures are fine if all you are protecting is \textquote{three fishbones and four sandwiches}.

Ultimately, we found that most providers in our study shared a similar philosophy, particularly those offering low-cost, standardized hosting services through public sign-up processes. In contrast, two providers primarily serving large businesses or public administration (\HPOtwelve{}, \HPOthirtn{}), as well as one with a mixed clientele (\HPOtwenty{}), demonstrated significantly greater attention to security. These providers described remediation processes that addressed more complex threats than typical WordPress compromises.

Our findings suggest that most hosting providers view customer instance compromise as posing minimal risk, considering proactive application-layer security systems as an optional enhancement rather than a necessity. Providers frequently targeted by low-sophistication attackers, such as ``script kiddies'', may be particularly likely to hold this view. In contrast, providers who implement comprehensive security measures tend to face more sophisticated threats and do not rely on standardized setups commonly targeted by automated attacks. This, in turn, may reinforce the disconnect observed with vulnerability notifications among the former group, as security in these hosting environments is regarded as an added benefit rather than a core requirement.

\begin{tcolorbox}[colframe=black, colback=white, boxrule=0.3pt, arc=0mm]
\paragraph{\textbf{Takeaways}} The security posture of hosting providers is shaped primarily by operational priorities and the need to protect their own infrastructure, rather than by concerns for individual customer environments. Resource monitoring is primarily intended to maintain service level agreements and manage legal risks, and any detection of abuse within customer spaces tends to be incidental rather than the result of targeted efforts. Providers generally express confidence in their technical safeguards to isolate issues and protect core systems. This aligns with and reinforces their passive stance towards application-level vulnerability notifications.
\end{tcolorbox}

\section{Discussion}
\label{sec:implications}
We conducted and analyzed semi-structured interviews to explore why hosting provider organizations often fail to remediate reported vulnerabilities. Our analysis focused on how vulnerability notifications are received and handled (\textbf{RQ1}), identifying the underlying criteria behind the limited remediation responses commonly noted by researchers (\textbf{RQ2}). 
This section summarizes our findings and future directions based on insights from our 24 participants.

\subsection{Comparison with Prior Works}
We investigated impactful factors highlighted by prior works, such as contact channels, email content, and features of the sender, as well as HPOs’ internal structures, analyzing whether and to what extent they contribute to the lack of VN remediation.

First, our study revisits the reachability issues observed in previous work~\cite{stock2016hey, cetin2017make, poteat2021you}. In fact, most hosting providers confirmed that their contact is available via WHOIS, for example through ARIN~\cite{arin} or RIPE~\cite{ripe} databases.
However, the issue of reachability persists when expanding the scope to actors not considered in prior studies, such as web agencies, who may have an actual impact on remediating reported issues.
Additionally, trust plays a different role than in earlier research, which emphasized general distrust~\cite{hennig2022standing, stock2018didn, maass2021effective}. Our participants noted that clear, technical emails with credible explanations are usually considered, though not always acted upon. 
Finally, while prior research emphasized inner-organizational technical and structural barriers to secure practices~\cite{huaman2021large, dietrich2018investigating, staddon2019s, alomar2020you}, and challenges with keeping systems up-to-date~\cite{li2019keepers, tiefenau2020security, jenkins2024not, martius2020does}, participants in our study reported no major roadblocks in managing their own infrastructure, suggesting that these factors are less critical for VN remediation~\cite{bondar2023internet}.
The observed lack of remediation was explained by hosting providers as a result of the nature of the vulnerability, often viewed as outside the provider's responsibility. This new framing also helps explain why operator awareness did not always lead to remediation~\cite{li2016you, stock2018didn, durumeric2014matter, bondar2023internet}.

Notably, the reasons reported by operators differ from those described in~\cite{ethembabaoglu2024unpatchables}, which investigates the lack of patching following vulnerability reporting in Dutch municipalities. While the concept of responsibility is central in both works, Ethembabaoglu et al. attribute non-patching to a lack of awareness leading to neglect~\cite{ethembabaoglu2024unpatchables}, whereas our study shows that remediation in commercial hosting providers is often refused based on contractual terms. In web agencies, remediation is only partially implemented, depending on business-related factors.

\subsection{Implications and Future Directions}

\subsubsection{VNs as a Source of Awareness}
We thoroughly investigated the technical infrastructure and security measures at hosting providers. Most providers believe their infrastructure is resilient to malicious activity occurring in customer environments. As a result, and sometimes also due to their large customer base, many providers do not closely monitor ongoing abuse or enforce remediation.
Previous works also observed the lack of proactive remediation by HPOs~\cite{canali2013role} which, complemented with website owners’ neglect or forgetfulness concerning the security posture of their websites~\cite{hellenthal2025usual, stover2023website}, underscores the relevance of vulnerability notification campaigns, which help fight internet abuse at scale, raising HPOs' awareness.

\subsubsection{Improving on Reachability}
The most straightforward, though not necessarily more effective, approach to increase remediation rates may be to notify those actually responsible. Identifying website owners at scale remains a well-known challenge~\cite{stock2018didn, maass2021effective}. To overcome this, researchers could explore more advanced automated methods for identifying contact points, selecting appropriate targets such as website owners or web agencies based on the issue type. 
Future work could involve AI agents capable of navigating websites~\cite{stafeev2024yurascanner}, locating contact information, and extracting relevant text. However, researchers should consider that many end users do not hire professionals to maintain their websites. The content of vulnerability notifications should therefore be tailored to the recipient, as prior studies have shown that end users interpret such messages very differently~\cite{hennig2023vision, hennig2022standing}.

At the same time, successful large-scale vulnerability notification has been shown to be feasible through paid services~\cite{Moura2024Characterizing}. This raises a broader question about whether the challenge lies in the lack of a standardized infrastructure available to researchers. Future work could address this by developing a shared contact database, similarly to the development of Tranco~\cite{le2018tranco} as an alternative to Alexa 1 Million, in order to reduce the impact of limited reachability on remediation outcomes.

\subsubsection{Lowering Cost of Remediation for Providers}
Many providers expressed a cynical view of the current state of abuse in shared hosting, where hackers compromise hundreds of sites with a single click and website owners are described as unskilled and careless, both in managing websites and in valuing website data. Especially among website hosting providers, the volume of daily abuse tickets was reported to be overwhelming (similarly to~\cite{de2023no}), leaving little room for careful attention to individual cases. A few providers mentioned occasionally resolving issues proactively, not out of policy but to avoid more costly customer service calls. Still, increasing staff to handle abuse reports more thoroughly appears impractical, as it would require raising prices in a highly cost-sensitive market.

Although providers set pretty strict responsibility boundaries, researchers could focus on making remediation easier and less resource-intensive for them. Many providers value technical detail, suggesting that notifications including clear evidence of the issue and detailed remediation steps could lead to higher response rates.

Moreover,  while simply updating the web application software is sometimes the most effective and straightforward remediation, this is often not feasible, as updates can cause incompatibilities with plugins. Our study shows that in such cases, providers often choose not to take action. Previous studies have shown that end users may be unaware of the need to update or may be unwilling to do so when updates impair website functionality~\cite{hellenthal2025usual}. This issue is further complicated by the open-source nature of many free web application tools, which are often untested or no longer maintained.
In such situations, researchers who discover the problem could help identify stable versions of the software to which updates are possible, or recommend alternative software that maintains overall functionality. End users, who often prioritize functionality over security, may otherwise never seek such alternatives, possibly due to a lack of awareness of the associated security risks.

\subsubsection{Beyond Reachability: Bridging Responsibility Gaps}
\new{While improving the reachability of vulnerability notifications remains an important step, our findings caution against treating it as a silver bullet. Simply reaching more providers does not guarantee remediation~\cite{stock2018didn, bondar2023internet}, as awareness alone rarely triggers remediation when responsibility is perceived to lie elsewhere. Instead, our study highlights a systemic gap: researchers expose risks, providers host the affected system yet reject ownership of the vulnerabilities, and end users--who are formally responsible--are often unaware, disinterested, or unmotivated to act~\cite{hellenthal2025usual}. Future work should focus on mechanisms that balance these competing positions, such as approaches that increase end-user awareness of downstream risks or tools that empower providers to intervene in compromised sites without alienating customers or incurring prohibitive costs.
Ultimately, improving remediation will require not just better messages, but a better alignment of incentives across all three types of actors.}

\subsection{Limitations}

The  field of research on vulnerability notification is broad, and our study does not aim to cover all possible scenarios. 
We focused specifically on web application vulnerabilities, whereas responses may differ for issues affecting HPO infrastructure, lower layers of the Internet stack, or internet-connected devices. 
We did not examine hosting services such as Dedicated Servers, Colocation, and Website Builders. These were excluded due to the nature of provider involvement in each case. Dedicated Servers and Colocation offer an extremely high degree of user control, with minimal provider intervention, while Website Builders represent the opposite extreme, offering limited customization through provider-specific, closed-source platforms.
\new{Our sample thus covers segments where providers retain some responsibility for the hosted software environment (shared hosting, VPS, and web agencies). While some VPS offerings may resemble dedicated-server models with little provider involvement, others involve substantial management by the provider.}
\new{As with most qualitative work, our findings are not statistically generalizable, e.g., to all parts of the hosting ecosystem, but aim for transferability by providing in-depth accounts across a diverse sample, reflecting structural variation in the hosting providers population studied. }

Secondly, we interviewed only one participant per HPO, potentially missing other perspectives on security in larger organizations. Despite our efforts, recruiting proved challenging. While interviewees provided valuable insights into company practices, future studies should conduct multiple interviews within the same HPOs to better capture innerorganizational dynamics of dealing with cy-sec issues and VN handling.
\new{Additionally, while we acknowledge that many providers in our sample are Europe-based, we remark that there was no such selection criteria on our side, and that this is merely a result of the choice of participants willing to respond to our messages.}

\new{We also acknowledge that our study relies on participants’ accounts rather than direct observation, and we did not experimentally validate providers’ claims. While we observed no signs of intentional withholding during interviews, some providers may not have been fully transparent about their practices for reputational reasons. Nonetheless, several participants openly discussed faulty systems or overlooked organizational procedures, suggesting a willingness to go beyond idealized accounts.}
Finally, our study may be influenced by opt-in as well as social desirability biases, and demand effects. To minimize these, we avoided disclosing the study’s exact focus on security and privacy topic during recruitment, encouraging open discussion and reducing bias related to prior VN experiences.


\section{Conclusion}
This study shifts the lens of vulnerability notification research by examining the recipients--hosting provider organizations--rather than optimizing the reporting process. To our best knowledge, this is the first study to deeply investigate how HPOs internally process and respond to vulnerability notifications. 
Through interviews with 24 HPOs, we find that while most providers are reachable and familiar with handling VNs, remediation remains limited due to structural and economic factors rather than technical ones.

Our findings indicate that low remediation rates are not primarily caused by unreachability, distrust, or internal procedural barriers. Instead, they stem from clear service boundaries, where responsibility for code vulnerabilities lies outside the scope of hosting providers’ obligations. Cost considerations, the commoditized nature of hosting services, and the perceived insignificance of individual customer environments further reduce the incentive to act. Moreover, remediation decisions are shaped by business priorities, ethical stances, and the discretion of individual operators rather than systematic security policies.

These insights challenge existing assumptions in VN literature and suggest that improving remediation rates requires addressing the underlying misalignment between the security goals of external reporters and the service logic of HPOs. Improving reachability remains important, especially by better targeting responsible parties such as website owners or web agencies. Yet the limited skills and engagement of many website owners complicate notification efforts, highlighting the need for messages tailored to non-experts.

\bibliographystyle{IEEEtran}
	\bibliography{50_bibliography} 

\appendix
{\renewcommand{\thesubsection}{\Alph{subsection}}

\subsection{Interview Guide}
\label{app:ig}
In the following, we present the interview questions used to conduct the semi-structured interviews. 

\subsubsection*{Introductory Questions}
\textbf{The company and interviewee's role in it.}
\begin{itemize}
    \item Please describe in as much detail as possible your role within your company.
    \item How is the provision of hosting services embedded into the general business
strategy of your company?
\end{itemize}

\subsubsection*{Technical Infrastructure Questions}
\textbf{General security posture of the company and identified risks.}
\begin{itemize}
    \item What are general security considerations that you deem relevant for the
operation of your company’s hosting services?
    \begin{itemize}
        \item What are possible types of attacks (`threat models') affecting hosting
    services provided by your company?
        \item What security-related risks are there for hosting services?
    \end{itemize}

    \item What does your company exactly do to prevent this specific kind of attack / threat?
    \begin{itemize}
        \item Are there specific departments in the company in charge of mitigating these risks?
        \item Are there in-house developed tools, or external tools?
    \end{itemize}
\end{itemize}

\textbf{Legal obligations for maintenance.}
\begin{itemize}
    \item If possible, can you describe the legal regulations that the company has to abide to, concerning security maintenance of the hosting services?
\end{itemize}

\subsubsection*{Organizational Aspects Questions}
\textbf{Security management.}
\begin{itemize}
    \item How is management/handling of security events organized in your company?
    \item Does your company manage security events (when they happen) internally (by its own staff), or are they outsourced to an external contractor?
        \begin{itemize}
        \item \textit{If handled externally:} Can you please describe which type of company is this and what services they offer?
        \item How are responsibilities distributed between your company (organization) and this external service provider?
    \end{itemize}
    
    \item Is there any set of (internal) instructions on what to do in such situations (a
``playbook'' of some kind)?
    \item Are there specific people (organizational units) who are in charge of
responding to security events?
    \begin{itemize}
        \item Does addressing the report involve multiple teams?
        \item Does the company distinguish between those who decide on remediation
    and those who implement it?
    \end{itemize}

    \item Does addressing the report involve external stakeholders?
\end{itemize}

\textbf{Personal security even handling experience.}
\begin{itemize}
    \item Did you ever experience any of these cybersecurity events at your company?
    \begin{itemize}
        \item How did your company learn about the (security) event?
        \item What was done in your company once the event was recognized?
        \item What was your specific role in this process?
        \item How did the whole situation end?
        \item What would you describe as ``lessons'' that your company learned from this event?
    \end{itemize}

    \item The course of action you described: Is it part of a procedure formalized by
the company, or is it something informal?
    \item Is this procedure regularly followed?
\end{itemize}

\subsubsection*{Vulnerability Notifications Questions}
\textbf{Being informed about risks and events.}
\begin{itemize}
    \item How are vulnerabilities and other S\&P problems detected in the infrastructure of your company’s hosting services?
    \item Does your company have a dedicated channel for reporting vulnerabilities and other S\&P problems?
    \item Does your organization receive vulnerability notifications from third parties?
    \item How does your company become aware of security events happening / happened?
    \begin{itemize}
        \item \textit{If no external source of info was mentioned:} Is there the possibility that the company would be informed of a security event from an external entity (person or organization) who were not contractors?
    \end{itemize}

\end{itemize}

\textbf{Vulnerability notification handling.}
\begin{itemize}
    \item \textit{If they have received VN:} What did you do with the external report?
    \item \textit{If they have never received a VN:} Imagine you‘ve just received a report from an external describing a security issue on your servers. 
    \item Please describe which factors  played/would play into your evaluation when deciding on how to act on this message.
	\item Which challenges do you face when evaluating external security reports on security events?
    \begin{itemize}
        \item How did/would the medium in which you received the report affect your decisions (e.g., email vs. contact form)?
		\item How did/would the identity of the sender of the report affect your decisions?
		\item What were/would be measures you applied/y to verify information in the report?
		\item How did/would legal requirements influence your decisions?
		\item How did/would your technical skills influence your decisions?
		\item How did/would authority-related questions influence your decisions?
		\item How did/would trust in the message in general affect your decisions?
    \end{itemize}
    \item Can you please walk us through the steps you take/took to remediate a security report?
    \item Have you ever received security reports about systems that were not on your perimeter?
\end{itemize}

\textbf{Vignettes describing specific VN issues.}
\begin{itemize}
    \item Malicious or suspicious behavior: a malware (e.g. \texttt{.exe}, or MS document with macro), a malicious file (e.g., a clickbait PDF), a file containing malicious code/functionality accessible from the web (e.g., a phishing page, a script redirecting to malicious/illegal content).
    \item Misconfiguration: a misconfiguration in HTTP headers enforcing security mechanisms (e.g. for transferring cookies), a misconfiguration in access control for a restricted area of a web app (e.g., default password, no access control).
    \item Code vulnerability: an error in the implementation of a software functionality, which can be observed via active testing or by passively observing indirect indicators (e.g., direct exploit of XSS, SQLi, or observation of software version indicators).

    \begin{itemize}
        \item How would you say your company considers these vulnerabilities?
        \item \textit{If they are considered:} Which tools does your company use to keep track of these vulnerabilities?
        \item \textit{If they are considered:} Is there any regular maintenance task to prevent them?
		\item \textit{If they are considered:} When deciding whether to address a vulnerability, would it influence your decision to know that it is a malicious file / misconfiguration / code vulnerability?
    \end{itemize}
\end{itemize}

\subsection{Codebook}
\label{app:codebook}

\subsubsection*{HPO}
\begin{itemize}
    \item Business Aspects
    \begin{itemize}
        \item Feel responsible for internet security
        \item Ideologically driven hosting policies
    \end{itemize}
    \item Business Considerations
    \item Services Offered
    \begin{itemize}
        \item Maintenance
    \end{itemize}
    \item Organizational Aspects
    \begin{itemize}
        \item Company Structure
        \begin{itemize}
            \item Communication Between Teams
            \item Departments
        \end{itemize}
        \item The infrastructure-customer divide
    \end{itemize}
    \item Functioning of Services
    \begin{itemize}
        \item Take Action to Ensure SLA
        \item Take Action to Safeguard Infrastructure Safety
        \item Manuals 
    \end{itemize}
    \item CySec Aspects
    \begin{itemize}
        \item Abuse of HPO Infrastructure
        \begin{itemize}
            \item Our infrastructure does not care about phishing
        \end{itemize}
        \item Security Posture
        \begin{itemize}
            \item General Proactive Measures		
            \item General Security Mindset		
            \begin{itemize}
                \item Keep updated	
            \end{itemize}
            \item Incident Response		
            \begin{itemize}
                \item Proactive-Security Practices	
                \item Proactive-Security Software
                \item Detection
                \item Detection Software
                \item Reactive-Security Practices
                \item Reactive-Security Software
                \item Incident Response
            \end{itemize}
            \item Organization Policies
            \item Weaknesses
            \begin{itemize}
                \item Customer Created
                \item HPO Created
                \item HPO-Customer Co-responsible
            \end{itemize}
        \end{itemize}
    \end{itemize}
    \item External Relations
    \begin{itemize}
        \item Externals Contacting HPO
        \begin{itemize}
            \item Brand/ Phishing Protection Companies
            \item Companies Reporting Phishing or Malware
        \end{itemize}
        \item Service Provider Relationship
    \end{itemize}
\end{itemize}

\subsubsection*{Customer}
\begin{itemize}
    \item Business model of your customer
    \item Characteristics
    \begin{itemize}
        \item Customer Structure
        \item Customer Tech-Savvyness
        \item Customer Types
    \end{itemize}
    \item Choice of Product / Service
    \item Customer Resources
    \begin{itemize}
        \item Own IT Team
        \item Own Software
    \end{itemize}
    \item Mindset of HPO Customers
    \item Reaction to Security Event
    \item Relationship \& Communication with HPO
    \begin{itemize}
        \item Customer Complains to HPO
        \item Customer Reports VN received
    \end{itemize}
    \item Security Posture of Customer
\end{itemize}

\subsubsection*{Events}
\begin{itemize}
    \item CySec Events
    \begin{itemize}
        \item How
        \item Detection Stage
        \item Post-Incident Stage
        \item Response Stage
    \end{itemize}
\end{itemize}

\subsubsection*{Participant}
\begin{itemize}
    \item Active Role
    \item Background / Professional Biography
\end{itemize}

\subsubsection*{Regulations}
\begin{itemize}
    \item Certification
    \item Documents with Legal Implication
    \item Must-Dos
    \item Regulators
    \item Won't-Dos
    \item Legal Regulations for Security
\end{itemize}

\subsubsection*{Security Third Parties}
\begin{itemize}
    \item CERT
    \item Google SafeBrowsing
    \item VirusTotal
\end{itemize}

\subsubsection*{Tech Used}
\begin{itemize}
    \item Server-Side Software
    \item Tech for Hosting
\end{itemize}

\subsubsection*{Vulnerability Notifications}
\begin{itemize}
    \item Assessment
    \begin{itemize}
        \item Criteria
        \begin{itemize}
            \item Business Criteria
            \item Interaction-Based Verification
            \item Validate Email Style
            \item Validate Email Metadata
            \item Validate Informative Content
        \end{itemize}
        \item Outcome
    \end{itemize}
    \item Challenges
    \begin{itemize}
        \item Customers Don't Understand
        \item Unclear Motivations of Senders
        \item Receive Spam
    \end{itemize}
    \item Communication Channels
    \item Handling Method
    \begin{itemize}
        \item Evaluation by Human
        \item Gut Feeling or Experience
    \end{itemize}
    \item Handling Practices
    \begin{itemize}
        \item Validation via Sender Communication
    \end{itemize}
    \item HPO's VN Knowledge
    \item Improvements
    \item Receiving Entity
    \begin{itemize}
        \item Company Receives Notification
        \item Customer Receives Notification
    \end{itemize}
    \item Sender
    \item VN Exposure
    \item VN reachability
\end{itemize}

\subsubsection*{Others}
\begin{itemize}
    \item Attack Types
    \item Information Sources
\end{itemize}

\newpage

\subsection{Meta-Review}

The following meta-review was prepared by the program committee for the 2026
IEEE Symposium on Security and Privacy (S\&P) as part of the review process as
detailed in the call for papers.

\subsubsection{Summary}
This paper presents a qualitative study of how 24 hosting providers manage vulnerability notifications, revealing that remediation efforts are shaped primarily by strict service boundaries, organizational structure, and the perceived responsibility for security issues, often resulting in limited action on reports. The findings highlight persistent gaps between the expectations of notification senders and real-world provider practices, emphasizing the need for clearer, more actionable recommendations for improving vulnerability remediation.

\subsubsection{Scientific Contributions}
\begin{itemize}
\item Provides a Valuable Step Forward in an Established Field
\end{itemize}

\subsubsection{Reasons for Acceptance}
\begin{enumerate}
\item Addresses a Key Gap: The work explores hosting providers’ operational and organizational factors influencing vulnerability notification effectiveness, a perspective not previously examined in depth. This helps surface reasons for persistently low remediation not explained by notification-centric studies.
\item Diverse, Relevant Observations: Through semi-structured interviews with representatives of 24 providers of varying types, the authors collect a rich set of observations, enabling critical insights into policy, responsibility boundaries, and impediments to remediation.
\item Enables Further Research: The authors plan to release an original dataset of the studied services, supporting transparency and laying the groundwork for future research.

\end{enumerate}

\subsection{Noteworthy Concerns}
\begin{enumerate} 
\item Limited Scope Reduces Generalizability: The study is heavily weighted toward European organizations and excludes dedicated server providers, which may limit applicability of the findings outside the sampled segments.
\item Depth of Analysis and Actionability: The paper's stated recommendations may not be specific enough to be actionable for either hosting providers or reporters. Proposals for report standardization or improved communication would benefit from clearer, more concrete guidance rooted in the interview data.
\item Potential Respondent Biases: The study relies on participants’ descriptions of their formal processes, which may differ from actual practices.
\end{enumerate}

}

\end{document}